\def\gtwid{\mathrel{\raise.3ex\hbox{$>$\kern-.75em\lower1ex\hbox{$\sim$}}}}
\def\ltwid{\mathrel{\raise.3ex\hbox{$<$\kern-.75em\lower1ex\hbox{$\sim$}}}}
\documentclass{elsart}

\usepackage{graphicx}
\usepackage{amsmath}
\usepackage{amssymb}

\begin{document}

\begin{frontmatter}

\title{Social Balance on Networks: The Dynamics of Friendship and Enmity}
\author{T.~Antal}
\author{P. L. Krapivsky}
\author{S.~Redner}
\address{Center for Polymer Studies and Department of Physics, Boston
  University, Boston, Massachusetts 02215}

\begin{abstract}
  
  How do social networks evolve when both friendly and unfriendly relations
  exist?  Here we propose a simple dynamics for social networks in which the
  sense of a relationship can change so as to eliminate imbalanced
  triads---relationship triangles that contains 1 or 3 unfriendly links.  In
  this dynamics, a friendly link changes to unfriendly or {\it vice versa} in
  an imbalanced triad to make the triad balanced.  Such networks undergo a
  dynamic phase transition from a steady state to ``utopia''---all friendly
  links---as the amount of network friendliness is changed.  Basic features
  of the long-time dynamics and the phase transition are discussed.

\end{abstract}

\begin {keyword}
Social balance \sep Networks

\PACS 02.50.Ey\sep
05.40.-a\sep
89.75.Fb
\end{keyword}
\end{frontmatter}

\section{Introduction}

As we all have experienced, social networks can evolve in convoluted ways.
Friendships can become estrangements and vice versa.  New friendships can be
created while existing friends drift apart.  How are these changing relations
reflected in the structure of social networks?  As a familiar and
illustrative example, suppose that you are friendly with a married couple
that gets divorced.  A dilemma arises if you try to remain friendly with both
of the former spouses.  You may find yourself in the uncomfortable position
of listening to each of the former spouses separately disparaging each other.
Ultimately you may find it simplest to remain friends with only one of the
former spouses and to cut relations with the other ex-spouse.  In the
language of {\em social balance} \cite{FH,Lewin,Newcomb,social}, the
initially balanced triad became unbalanced when the couple divorced.  When
you subsequently kept your friendship with only one former spouse, social
balance is restored.

\begin{figure}[ht]
\centerline{  \includegraphics[width=0.8\linewidth]{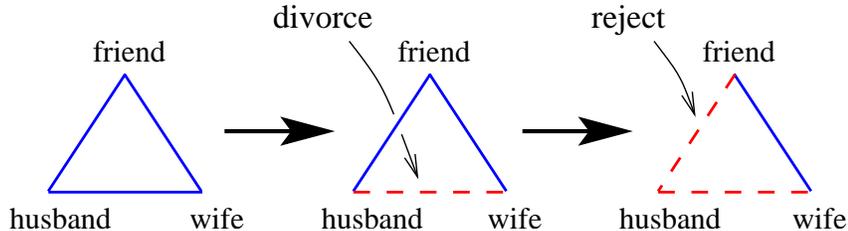}}
  \caption{Evolution of a married couple plus friend triad.  After a divorce
    the triad becomes imbalanced, but balance is restored after another
    relationship change.  Full and dashed lines represent friendly and
    unfriendly relations respectively.  }
\label{triads}
\end{figure}

What happens in a larger social network?  Now we need to look at all triads
$ijk$ that link individuals $i$, $j$, and $k$.  We define the link variable
$s_{ij}=1$ if $i$ and $j$ friends and $s_{ij}=-1$ otherwise.  Then the triad
$ijk$ is balanced if $s_{ij}s_{jk}s_{ki}=1$, and is imbalanced otherwise
(Fig.~\ref{triads}).  A balanced triad therefore fulfills the adage:
\begin{itemize}
\item a friend of my friend as well as an enemy of my enemy is my friend;
\item a friend of my enemy as well as an enemy of my friend is my enemy.
\end{itemize}
A network is balanced if each constituent triad is balanced \cite{FH,social}.
A seemingly more general definition of a balanced network is to require that
each closed cycle is balanced; that is, $\prod_{\ell\in {\rm path}}
s_{\ell}=+1$.  Cartwright and Harary showed \cite{CH} that a cycle-based
definition of balance is equivalent to a triad-based definition for complete
graphs.  This result can be reformulated as follows: if we detect an
imbalanced cycle of any length in a complete graph, there must be an
imbalanced triad.

Balance theory was originally introduced by Heider \cite{FH} and important
contributions were made by many others \cite{Lewin,Newcomb,New}.  Cartwright
and Harary \cite{CH,HNC} translated Heider's ideas into the framework of
graph theory, and proved several fundamental theorems about the structure
of balanced networks.  There is also an extensive literature on balance
theory (see {\it e.g.}, \cite{social,Leik,B,Davis,Hum,Dor,HD} and references
therein).

Cartwright and Harary showed that on a complete graph balanced societies are
remarkably simple: either all individuals are mutual friends (``utopia''), or
the network segregates into two mutually antagonistic but internally friendly
cliques---a ``bipolar'' state \cite{CH}.  However, spontaneously balanced
states are rare---if one were to assign relationships in a social network at
random, the probability that this society is balanced would vanish
exponentially with system size.  Thus to understand how a network reaches a
balanced state we need to go beyond static descriptions to investigate how an
initially imbalanced society becomes balanced via social dynamics.

Here we discuss the evolution of such social networks when we allow the sense
of each link to change from friendly to unfriendly or {\it vice versa} to
reflect the natural human tendency to reduce imbalanced triads \cite{AKR,K}.
Two such dynamics are considered: {\em local triad dynamics} (LTD) and {\em
  constrained triad dynamics} (CTD).  For simplicity, we consider complete
graph networks---everyone knows everyone else.  We will address the basic
question: what is the long-time state of such networks?

\section{Local Triad Dynamics}

\subsection{The Update Rule}

In local triad dynamics (LTD), an imbalanced triad is selected at random and
the sign of a relationship between two individuals is flipped to restore the
triad to balance.  This change is made irregardless if other triads become
imbalanced as a result.  Thus LTD can be viewed as the social graces of the
clueless---such a person makes a relationship change without considering the
ramifications on the rest of his social network.  We define a triad
$\triangle$ to be of type $k$ if it contains $k$ unfriendly links.  Thus
$\triangle_0$ and $\triangle_2$ are balanced, while $\triangle_1$ and
$\triangle_3$ are imbalanced.  With these definitions, the LTD rules are
(Fig.~\ref{process}):

\begin{figure}[ht]
\centerline{  \includegraphics[width=0.8\linewidth]{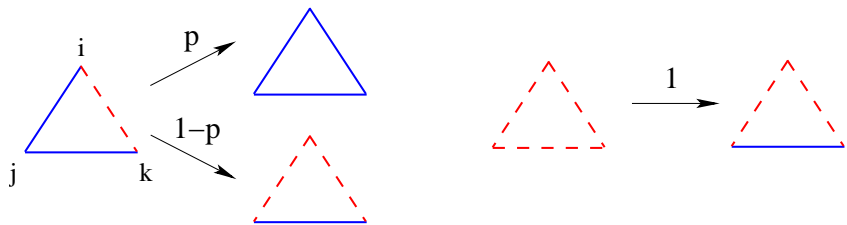}}
  \caption{An update step on imbalanced triads $\triangle_1$ (left) and
    $\triangle_3$ (right) by local triad dynamics.  Solid and dashed lines
    represent friendly and unfriendly links, respectively.}
\label{process}
\end{figure}

\begin{enumerate}
\item Pick a random imbalanced (frustrated) triad.
\item If the triad is of type $\triangle_1$, then: (i)
with probability $p$, change the unfriendly link to a friendly link; (ii)
with probability $1-p$, change a friendly link to an unfriendly link.
\item If the triad is of type $\triangle_3$, then change an unfriendly link
  to a friendly link.
\end{enumerate}

After the update, the initial imbalanced target triad becomes balanced, but
other previously-balanced triads that share a link with this target may
become imbalanced.  These triads can subsequently evolve and return to
balance, leading to new imbalanced triads.  For example, when a married
couple breaks up, friends of the former couple that remain friends with the
former wife may then redefine their relationships with those who choose to
remain friends with the former husband.  These redefinitions, may lead to
additional relationship shifts, {\it etc}.

\subsection{Evolution on the Complete Graph}

We now study LTD on a finite complete graph of $N$ nodes, $L=\binom{N}{2}$
links, and $N_{\triangle} = \binom{N}{3}$ triads.  Let $N_k$ be the number of
triads that contain $k$ unfriendly links, with $n_k=N_k/N_\triangle$ the
respective triad densities, and $L^+$ ($L^-$) the number of friendly
(unfriendly) links.  The number of triads and links are related by
\begin{equation}
L^+ \!=\! \frac{3N_0+2N_1+N_2}{N-2}\,,\quad
L^-\!=\!\frac{N_1+2N_2+3N_3}{N-2}\,\,.
\label{L+-}
\end{equation}
The numerator counts the number of friendly links in all triads while the
denominator appears because each link is counted $N-2$ times.  The density of
friendly links is therefore $\rho=L^+/L=(3n_0+2n_1+n_2)/3$, while the density
of unfriendly links is $1-\rho=L^-/L$.

\begin{figure}[ht]
\centerline{   \includegraphics[width=0.6\linewidth]{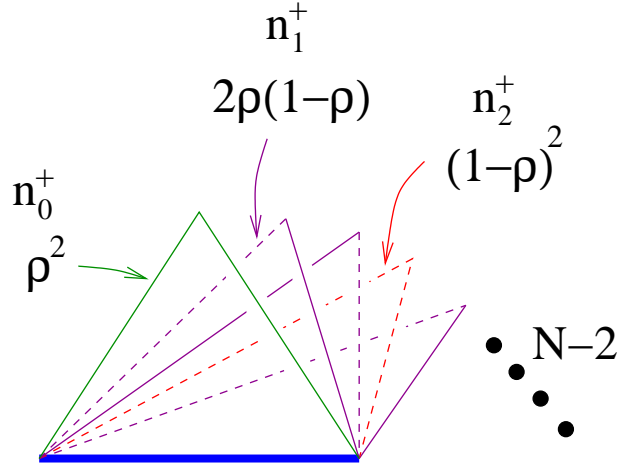}}
  \caption{Illustration of the different types of triads ($N-2$ in total)
    that are attached to a positive link (heavy line).  Also shown are the
    stationary-state probabilities for each triad when the friendly link
    density is $\rho$.  Full and dashed lines represent friendly and
    unfriendly relations, respectively.}
\label{triangles-plus}
\end{figure}

It is useful to introduce the quantities $N_k^+$ as follows: for each
friendly link, count the number of triads of type $\triangle_k$ that are
attached to this link.  Then $N_k^+$ is the average number of such triads
over all friendly links.  This number is
\begin{equation}
\label{Ni+}
N_k^+ = \frac{(3-k)N_k}{L^+}.
\end{equation}
The factor $(3-k)N_k$ accounts for the fact that each of the $N_k$ triads of
type $\triangle_k$ is attached to $3-k$ friendly links; dividing by $L^+$
then gives the average number of such triads.  Analogously, we introduce
$N_k^-=kN_k/L^-$.  Since the total number of triads attached to any given
link equals $N-2$, the corresponding triad densities are
(Fig.~\ref{triangles-plus})
\begin{subequations}
\begin{align}
&n_k^+=\frac{N_k^+}{N-2} = \frac{(3-k)n_k}{3n_0+2n_1+n_2}
\label{ni+}\\
&n_k^-=\frac{N_k^-}{N-2} = \frac{kn_k}{n_1+2n_2+3n_3} ~.
\label{ni-}
\end{align}
\end{subequations}

We now write rate equations that account for the changes in the triad densities
in an update.  We choose a triad at random; if it is imbalanced
($\triangle_1$ or $\triangle_3$) we change one of its links as shown in
Fig.~\ref{process}.  Let $\pi^+$ be the probability that a link changes from
friendly to unfriendly in an update event, and vice versa for $\pi^-$.  A
friendly link changes to unfriendly with probability $1-p$ when
$\triangle_1\to \triangle_2$, while an unfriendly link changes to friendly
with probability $p$ if $\triangle_1\to \triangle_0$ and with probability 1
if $\triangle_3\to \triangle_2$.  Consequently
\begin{equation}
\label{pi}
\pi^+ = (1-p)\,n_1 \qquad \pi^- = p\,n_1+n_3.
\end{equation}
In the special case of $p=1/3$, each link of an imbalanced triad is flipped
equiprobably.  Since each update changes $N-2$ triads, and we define one time
step as $L$ update events.  Then the rate equations for the triad densities
have the size-independent form
\begin{equation}
\label{ni-rate}
\begin{split}
&\dot n_0  =  \pi^- n_1^- - \pi^+ n_0^+\,,\\ 
&\dot n_1  =  \pi^+ n_0^+ + \pi^- n_2^- - \pi^- n_1^- - \pi^+n_1^+\,,\\
&\dot n_2  =  \pi^+ n_1^+ + \pi^- n_3^- - \pi^- n_2^- - \pi^+n_2^+\,,\\
&\dot n_3  =  \pi^+ n_2^+ - \pi^- n_3^-\,,
\end{split}
\end{equation}
where the overdot denotes the time derivative.

\begin{figure}[ht]
\centerline{   \includegraphics[width=1.0\linewidth]{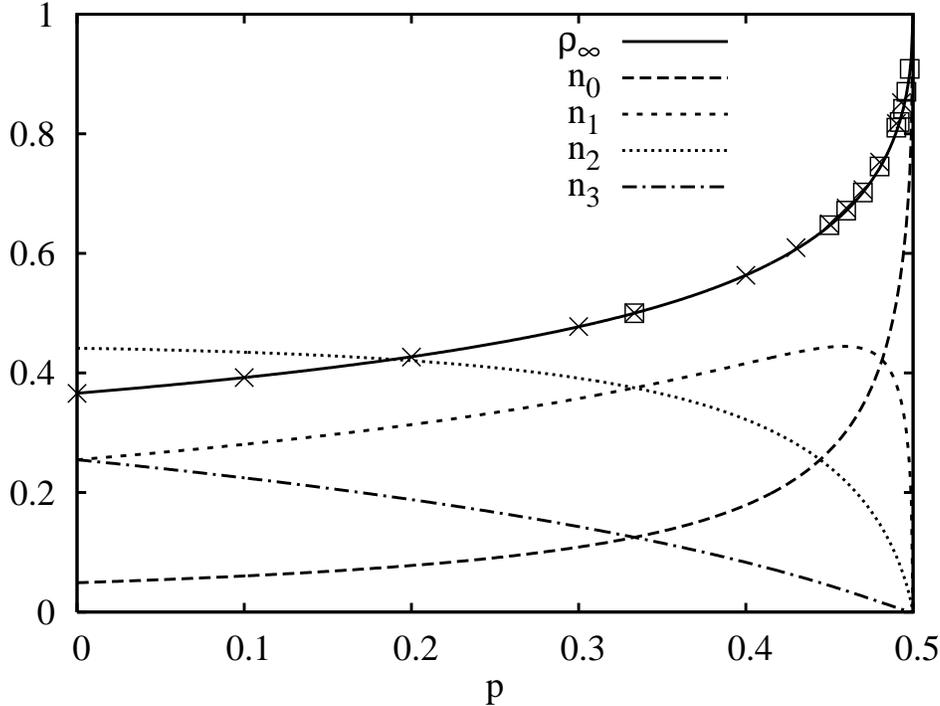}}
  \caption{The stationary densities $n_k(p)$ and the density of friendly
    links $\rho_\infty$ as a function of $p$.  Simulation results for
    $\rho_\infty$ for $N=64$ (crosses) and 256 (boxes) are also shown.}
\label{n-vs-p}
\end{figure}

Let us determine the stationary solution to these equations.  Setting the
left-hand sides of Eqs.~(\ref{ni-rate}) to zero and also imposing
$\pi^+=\pi^-$ to ensure a fixed friendship density, we obtain $n_0^+ =
n_1^-~,~~ n_1^+ = n_2^-~,~~ n_2^+=n_3^-$.  Forming products such as
$n_0^+n_2^-=n_1^+n_1^-$, these relations are equivalent to
\begin{equation}
\label{stati} 
3n_0 n_2 = n_1^2~,\qquad 3n_1 n_3 = n_2^2\,. 
\end{equation}
Furthermore, the stationarity condition, $\pi^+=\pi^-$, gives
$n_3=(1-2p)n_1$.  Using these two results, as well as the normalization
condition, $\sum n_k=1$, in Eqs.~(\ref{stati}), we find, after straightforward
algebra, that the stationary density of friendly links is
\begin{equation}
\label{stat-friends}
\rho_\infty=
\begin{cases}
1/[\sqrt{3(1-2p)}+1]   &\qquad    p\leq 1/2;\cr       
1                      &\qquad    p\geq 1/2.\cr
\end{cases}
\end{equation}
The triad densities of each type become uncorrelated and are given by
\begin{equation}
\label{stat-nj}
n_j=\binom{3}{j} \rho_\infty^{3-j}(1-\rho_\infty)^{j}.
\end{equation}
As shown in Fig.~\ref{n-vs-p}, the stationary density of friendly links
$\rho_\infty$ monotonically increases with $p$ for $0\leq p\leq 1/2$ until
utopia is reached.  Near the phase transition, the density of unfriendly
links $u\equiv 1-\rho_\infty$ vanishes as $\sqrt{3(1-2p)}$.

\subsection{The Evolving State} 
\label{evol} 

A remarkable feature of the master equations (\ref{ni-rate}) is that if the
initial triad densities are given by Eq.~(\ref{stat-nj})---uncorrelated
densities---the densities will remain uncorrelated forever.  In this case, it
suffices to study the time evolution of the density of friendly links
$\rho(t)$.  We determine this time evolution directly by noting that
$\rho(t)$ increases if $\triangle_3\to \triangle_2$ or $\triangle_1\to
\triangle_0$, and decreases if $\triangle_1\to \triangle_2$.  Since the
respective probabilities for these processes are $1, p$, and $1-p$, we have
\begin{equation}
\label{rho}
\frac{d\rho}{dt}=3(2p-1)\rho^2(1-\rho)+(1-\rho)^3.
\end{equation}
Solving this equation, the time dependence of the density of friendly links
has the following behaviors:
\begin{equation}
\label{rho-cases}
\rho(t)-\rho_\infty\sim
\begin{cases}
{\displaystyle A e^{-Bt}} & p<1/2\\ \\
{\displaystyle -\frac{1-\rho_0}{\sqrt{1+2(1-\rho_0)^2t}}} & p=1/2 \\ \\
{\displaystyle - Ce^{-3(2p-1)t}}  &p>1/2\,,
\end{cases}
\end{equation}
where $A$, $B$, and $C$ are constants.  Thus for $p\ne 1/2$ there is quick
approach to a final state.  This state is frustrated for $p<1/2$ and is
utopian for $p>1/2$.  For $p=1/2$ utopia is reached slowly---as a power-law
in time.

\subsection{Fate of a Finite Society} 
\label{fate} 

Abstractly, LTD represents a stochastic dynamics in a state space in which
each network configuration is represented by a point in this space and a link
to another point represents an allowed transition by the dynamics.  Because
balanced networks represent absorbing states of this dynamics, a finite
network must ultimately fall into a balanced state for all $p$.  We now
estimate the size dependence of the time to reach a balanced state, $T_N$,
for any value of $p$ by probabilistic arguments.

\begin{figure}[ht]
\centerline{   \includegraphics[width=0.8\linewidth]{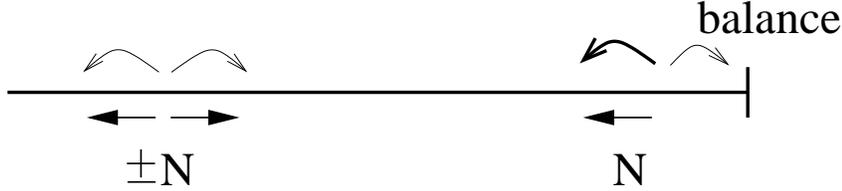}}
  \caption{Effective random walk picture for the approach to balance for
    $p<1/2$.}
\label{eff-rw}
\end{figure}

For $p<1/2$, we use the following random walk argument (Fig.~\ref{eff-rw}):
when a link is flipped on an imbalanced triad on an almost balanced network
(nearly $N^3/6$ balanced triads), then of the order of $N$ triads that
contain this link will become imbalanced.  Thus starting near balance, LTD is
equivalent to a biased random walk in the state space of all network
configurations, with the bias is directed away from balance, and with the
bias velocity $v$ proportional to $N$.  Conversely, far from the balanced
state, local triad dynamics is diffusive because the number of imbalanced
triads changes by of the order of $\pm N$ equiprobably in a single update.
The corresponding diffusion coefficient $D$ is then proportional to $N^2$.
Since the total number of triads in a network of $N$ nodes is
$N_\triangle\sim N^3/6$, we therefore expect that the time $T_N$ to reach
balance will scale as $T_N\sim e^{v N_\triangle/D}\sim e^{N^2}$ \cite{fpp}.

\begin{figure}[htb]
\centerline{      \includegraphics[width=0.7\linewidth]{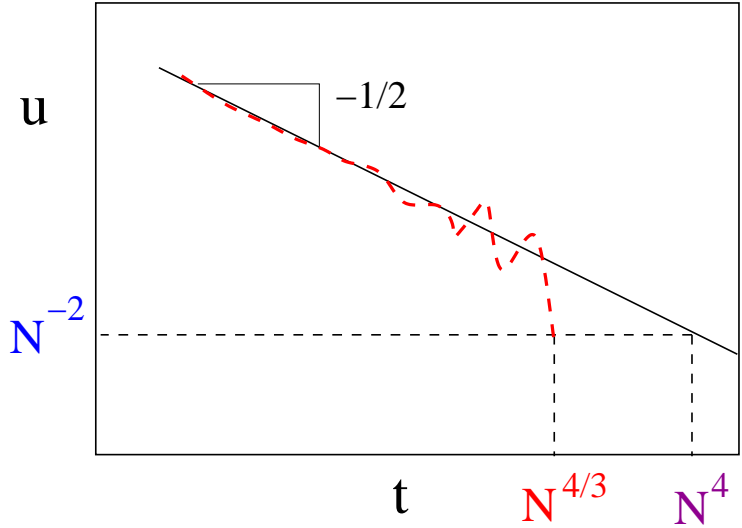} }
\caption{Illustration of the rate equation solution for the unfriendly link
  density versus time on a double logarithmic scale and the influence of
  fluctuations on this solution.  }
\label{rate-eqn}
\end{figure}

For $p> 1/2$, we define the time to reach the balanced state by the naive
criterion $u(t)\equiv 1-\rho(t)=N^{-2}$; that is, one unfriendly link
remains.  From Eq.~\eqref{rho-cases}, $T_N$ will then grow logarithmically
with $N$.  At $p=1/2$, using Eq.~(\ref{rho-cases}), the criterion
$u(t)=N^{-2}$ now gives $T_N\sim N^4$.  While simulations show that $T_N$
scales algebraically with $N$, the exponent is much smaller than 4.  The
source of this smaller exponent is the fact that the number of unfriendly
links fluctuates strongly about its mean value when there are few unfriendly
links (see Fig.~\ref{rate-eqn}).  To determine these fluctuations we write
the number of unfriendly links in the canonical form \cite{vK}
\begin{equation}
\label{expansion}
U(t)=Lu(t)+\sqrt{L}\,\eta(t),
\end{equation}
where $u(t)$ is deterministic and $\eta(t)$ is a stochastic variable.  Both
$u$ and $\eta$ are size independent in the thermodynamic limit.  A detailed
argument \cite{AKR} shows that $\sigma\equiv \langle\eta^2\rangle$ grows as
$\sigma\sim \sqrt{t}$ as $t\to\infty$.  Because of the finite-size
fluctuations in $U$, the time to reach utopia $T_N$ is determined by the
criterion that fluctuations in $U$ become of the same order as the average,
{\it viz.},
\begin{equation}
\label{criterion}
 \sqrt{L\sigma(T_N)}\sim L u(T_N) ~.
\end{equation}
Using $u(t)\sim 1/\sqrt{t}$ from Eq.~(\ref{rho-cases}), $\sigma\sim
\sqrt{t}$, and $L\sim N^2$, Eq.~(\ref{criterion}) becomes $ N\, T_N^{1/4}
\sim N^2\, T_N^{-1/2}$, from which $T_N$ follows.

\begin{figure}[htb]
\centerline{      \includegraphics[width=1.0\linewidth]{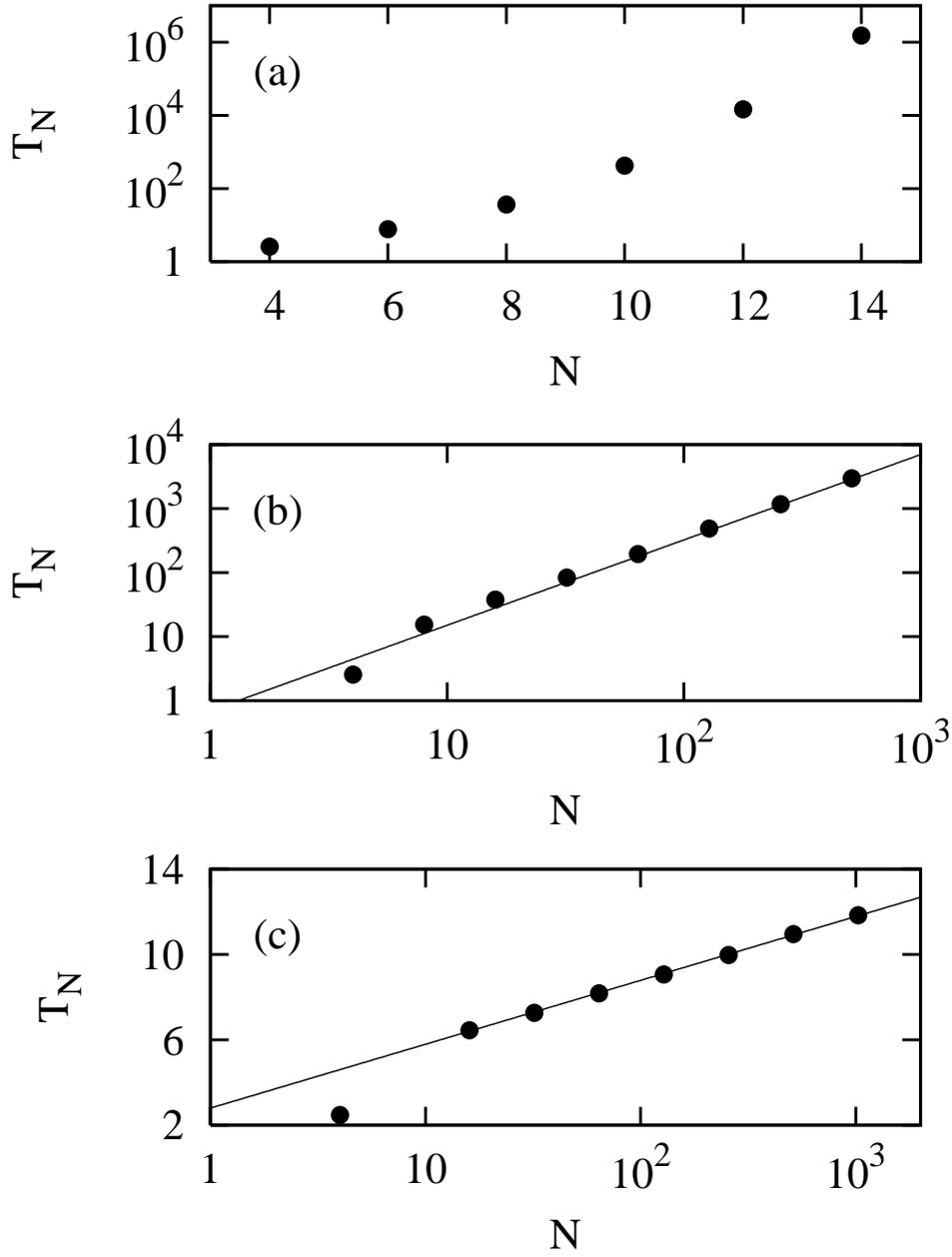} }
\caption{Average time to reach balance as a function of $N$ for an initially
  antagonistic society ($\rho_0=0$) for: (a) $p=1/3$; (b) $p=1/2$; (c) p=3/4.
  The line in (b) has slope $4/3$. }
\label{avtime}
\end{figure}

Summarizing our results, we have:
\begin{equation}
\label{BM-T}
T_N\propto 
\begin{cases}
e^{N^2}        &p<1/2\\
N^{4/3}                     &p=1/2\\
(2p-1)^{-1}\,\ln N          &p>1/2\,.
\end{cases}
\end{equation}
These are in agreement with our simulation results shown in
Fig.~\ref{avtime}.

\section{Constrained Triad Dynamics}

In {\em constrained triad dynamics} (CTD), we first select an imbalanced
triad at random and then select a {\em random} link in this triad.  We change
the sign of the link {\em only if the total number of imbalanced triads
  decreases}.  If the total number of imbalanced triads is conserved in an
update, then the update occurs with probability 1/2.  CTD can be viewed as
the dynamics of a socially aware individual who considers her entire social
circle before making any relationship change.  Because of this global
constraint, a network is quickly driven to a balanced state in a time that
scales as $\ln N$.

\begin{figure}[htb]
\centerline{      \includegraphics[width=1.0\linewidth]{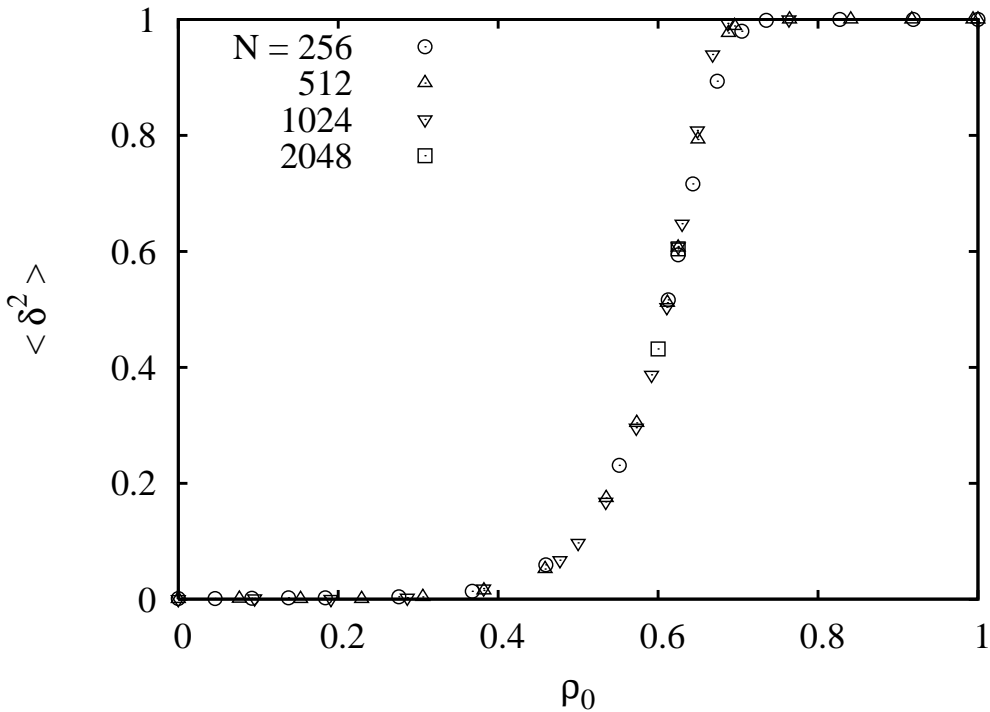} }
     \caption{Asymmetry of the final state as a function of the
       initial friendship density $\rho_0$ for several network sizes.}
\label{BavDi}
\end{figure}

A more interesting feature is the existence of a dynamical phase transition
in the structure of the final state as a function of the initial friendly
link density $\rho_0$ (Fig.~\ref{BavDi}).  We quantify this structural change
by the scaled difference in sizes of the two cliques in the final state,
$\delta\equiv (C_1-C_2)/N$.  For $\rho_0< 0.4$ the cliques in the final state
are nearly the same size and $\langle\delta^2\rangle\approx 0$.  As $\rho_0$
increases toward $\rho_0^* \approx 2/3$, the size difference continuously
increases and a sudden change occurs at $\rho_0^*$, beyond which the final
state is utopia.  Since $\langle \delta^2\rangle$ and the density of friendly
links $\rho_\infty$ are related by $\langle \delta^2 \rangle =
2\rho_\infty-1$ in a large balanced society, uncorrelated initial relations
generically lead to $\rho_\infty>\rho_0$.  Thus CTD tends to drive a network
into a friendlier final state.

We now give a simple-minded argument that suggests that a large network
undergoes a sudden change from $\rho_\infty=0$ (two equal size cliques) to
$\rho_\infty=1$ (utopia) as a function of the initial friendly link density
$\rho_0$.  This qualitative approach predicts that this transition occurs at
$\rho_0=1/2$.  On the other hand, our numerical simulations show that the
transition is located near $\rho^*_0\approx 2/3$ (Fig.~\ref{BavDi}).  

Let us assume that a network remains uncorrelated during initial stages of
evolution and under this assumption we determine the probabilities for a
specific friendly link to flip.  If the network is uncorrelated, the
densities ${\bf n^+}\equiv (n_0^+,n_1^+,n_2^+,n_3^+)$ of triads that are
attached to a friendly link are:
\begin{equation}
\label{positive} 
{\bf n^+}=(\rho^2,2\rho(1-\rho),(1-\rho)^2,0).
\end{equation}
For a link to change from friendly to unfriendly, it is necessary that
$n_1^++n_3^+ > n_0^++n_2^+$.  That is, this link is a member of more
imbalanced triads than balanced triads.  From Eq.~(\ref{positive}), this
condition is equivalent to $4\rho(1-\rho)>1$, which never holds.
Consequently, friendly links never flip.  Similarly, the densities ${\bf
  n^-}\equiv(n_0^-,n_1^-,n_2^-,n_3^-)$ of triads attached to an unfriendly
link are:
\begin{equation}
\label{negative} 
{\bf n^-}=(0,\rho^2,2\rho(1-\rho),(1-\rho)^2).
\end{equation}
To flip this unfriendly bond, we must have $n_1^-+n_3^->n_0^-+n_2^-$, {\it
  i.e.}, the bond is part of more imbalanced than balanced triads.  This
condition gives $1>4\rho(1-\rho)$, which is valid when $\rho\ne 1/2$.  Thus
for a large uncorrelated network, only unfriendly links flip in CTD, except
for $p=1/2$.  Thus a network with $\rho_0>1/2$ should quickly evolve to
utopia, while a network with $\rho_0<1/2$ should quickly approach a state
where $\rho=1/2$.

Simulations indicate, however, that correlations in relationships occur when
$\rho\approx 1/2$ and these ultimately lead to a bipolar society.  We find
that the precursor to this bipolar society is a state in which the network
partitions itself by the dynamics into two subnetworks ${S}_1$ and ${S}_2$ of
nearly equal sizes $C_1=|{S}_1|$ and $C_2=|{S}_2|$.  Within each subnetwork,
the density of friendly links $\rho_1$ and $\rho_2$ slightly exceeds $1/2$,
while the density $\beta$ of friendly links between subnetworks is slightly
less than $1/2$.  This small fluctuation is amplified by CTD so that the
final state is two nearly equal-size cliques.

To see how such evolution occurs, let us assume that relationships within
each subnetwork and between subnetworks are homogeneous.  Consider first the
evolution within each clique.  For an unfriendly link in ${S}_1$, the
densities of triads attached to this link are given by (\ref{negative}), with
$\rho$ replaced by $\beta$ when the third vertex in the triad belongs to
${S}_2$, and by (\ref{negative}), with $\rho$ replaced by $\rho_1$ when the
third vertex belongs to ${S}_1$.  The requirement that a link can change from
unfriendly to friendly by CTD now becomes
\begin{equation}
\label{req1} 
C_1[1-4\rho_1(1-\rho_1)]+C_2[1-4\beta(1-\beta)]>0,
\end{equation}
which is always satisfied.  Conversely, friendly links within each subnetwork
can never change.  As a result, negative intraclique links disappear and
there is increased cohesiveness within cliques.

\begin{figure}[htb]
\centerline{      \includegraphics[width=0.8\linewidth]{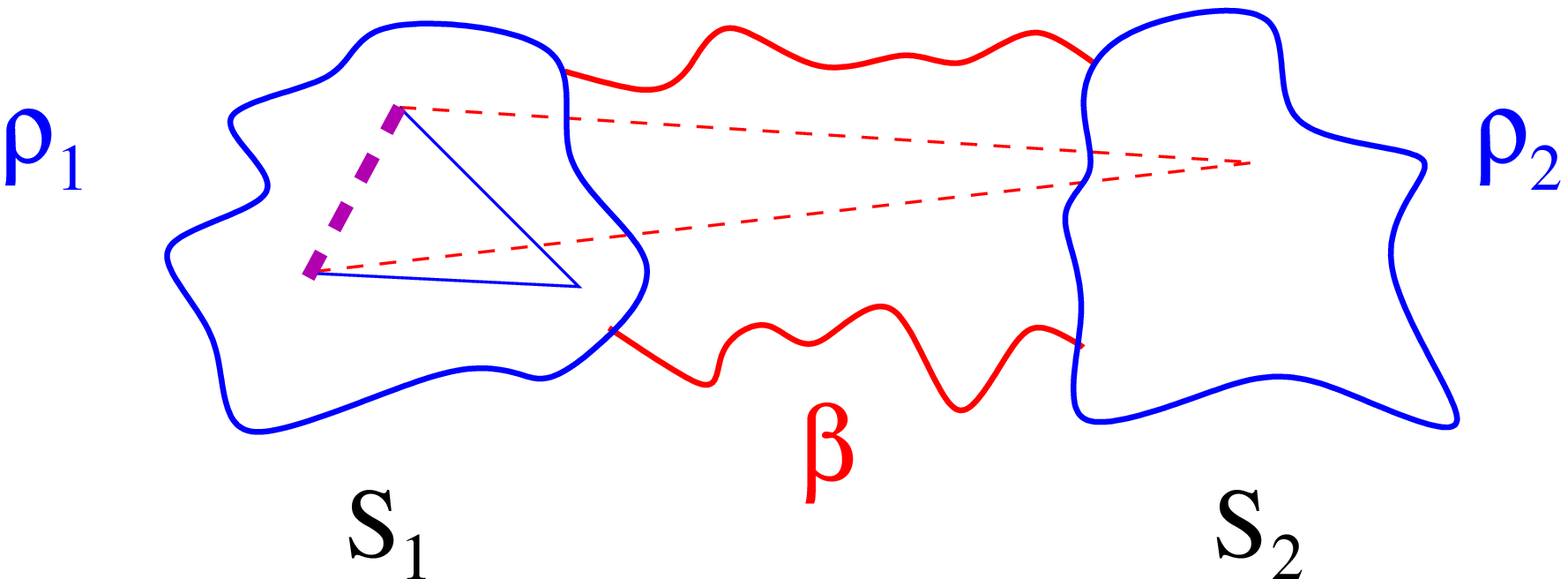}}
 \vskip 0.2in
 \centerline{     \includegraphics[width=0.68\linewidth]{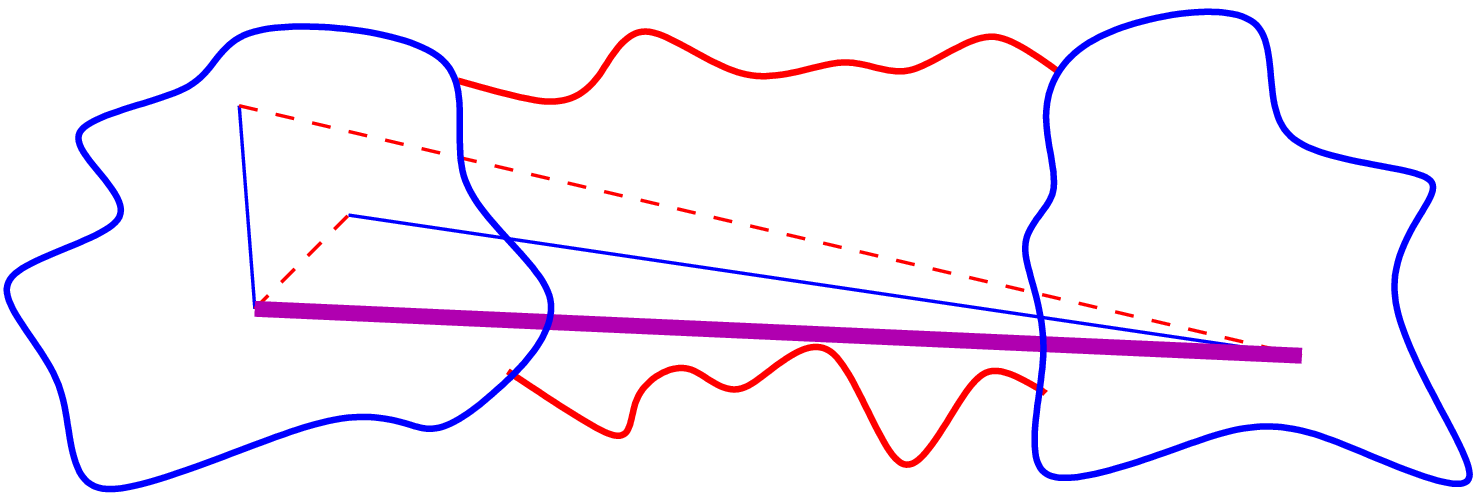} }
     \caption{Nascent cliques $S_1$ and $S_2$ (blobs at the extremities),
       with friendly link densities $\rho_1, \rho_2 \gtwid \frac{1}{2}$.  The
       density of friendly links between cliques is $\beta \ltwid
       \frac{1}{2}$.  Top: imbalanced triads that lead to an unfriendly link
       (think dashed line) changing to a friendly link within one clique.
       Bottom: imbalanced triads that lead to a friendly link (thick solid
       line) changing to a unfriendly link between cliques.}
\label{inter-intra}
\end{figure}

Consider now relations between cliques.  For a friendly link between the
subnetworks, the triad densities attached to this link are
\begin{equation*} 
{\bf n}^+_j=(\beta\rho_j,\beta(1-\rho_j)+\rho_j(1-\beta),
(1-\beta)(1-\rho_j),0)
\end{equation*}
when the third vertex belongs to ${S}_j$.  Since
\begin{equation*}
  \beta(1-\rho_j)+\rho_j(1-\beta)-\beta\rho_j-(1-\beta)(1-\rho_j)=
  (2\rho_j-1)(1-2\beta)\,,
\end{equation*}
the change friendly $\to$ unfriendly is possible if
\begin{equation}
\label{req2} 
[C_1(2\rho_1-1)+C_2(2\rho_2-1)](1-2\beta)>0\,.
\end{equation}
Thus if the situation arises where $\rho_1>1/2$, $\rho_2>1/2$, and
$\beta<1/2$, the network subsequently evolves to increase the density of
intra-subnetwork friendly links and decrease the density of inter-subnetwork
friendly links.  This bias drives the network to a final bipolar state.

Finally, note that when $C_1\approx C_2\approx N/2$, the number of ways,
$N\choose C_1$, to partition the original network into the two nascent
subnetworks ${S}_1$ and ${S}_2$, is maximal.  Consequently, the partition in
which $C_1=C_2$ has the highest likelihood of providing the initial link
density fluctuation that ultimately leads to two nearly equal-size cliques,
as observed in our simulations (Fig.~\ref{BavDi}).  Although our argument
fails to account for the precise location of the transition, the behavior of
$\langle \delta^2\rangle$ in the two limiting cases of $\rho_0\to 0$ and
$\rho_0\to 1$ is described correctly.

\section{Summary and Discussion}

We presented a simple setting for social dynamics in which both friendly and
unfriendly links exist in a network.  These links evolve according to natural
rules that reflect a social desire to avoid imbalanced triads.  For local
triad dynamics, a finite network falls into a socially-balanced state in a
time that depends sensitively on the propensity $p$ for forming a friendly
link in an update event.  For an infinite network, a balanced state is never
reached when $p<1/2$ and the system remains stationary.  The density of
unfriendly links gradually decreases and the network undergoes a dynamical
phase transition to an absorbing, utopia state for $p\geq 1/2$.  

For constrained triad dynamics, an arbitrary network is quickly driven to a
balanced state.  This rapid evolution results from the condition that the
number of imbalanced triads cannot increase.  There is also a phase
transition from bipolarity to utopia as a function of the initial density of
friendly links that arises because of small structural fluctuations that are
then amplified by the dynamics.

\begin{figure}[htb]
\centerline{      \includegraphics[width=1.0\linewidth]{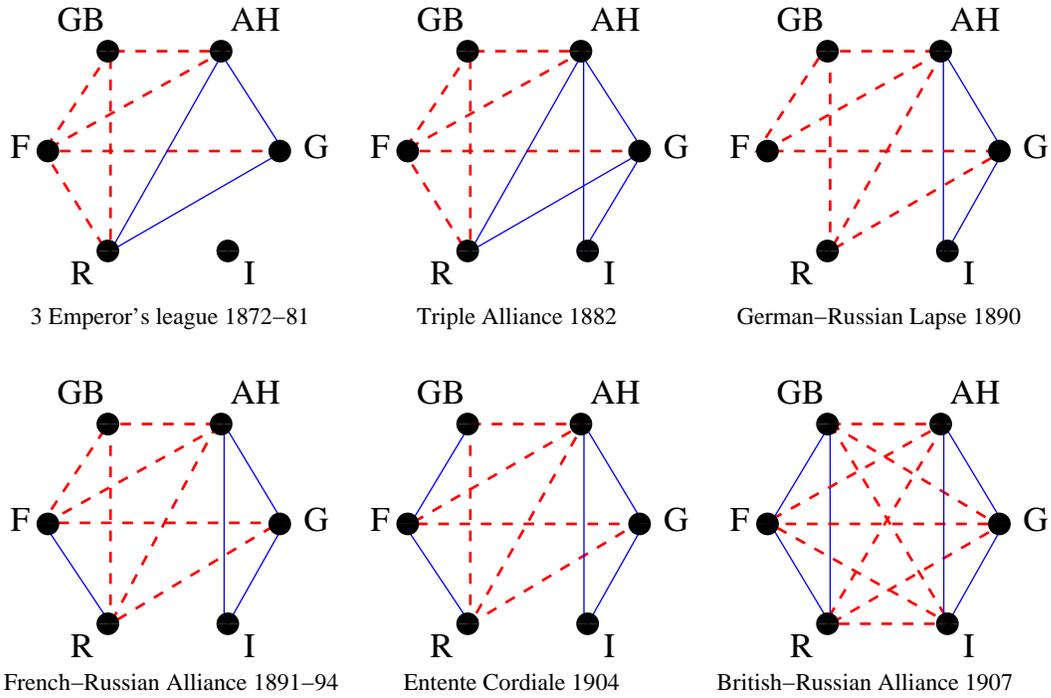} }
     \caption{Evolution of the major relationship changes between the
       protagonists of World War I from 1872--1907.  Here GB = Great Britain,
       AH = Austria-Hungary, G = Germany, I = Italy, R = Russia, and F =
       France.  }
\label{ww1}
\end{figure}

It is interesting to consider the possible role of balance theory in
international relations \cite{Moore}, with the evolution of the relations
among the protagonists of World War I being a particularly compelling example
(Fig.~\ref{ww1}).  A history starts with the Three Emperors' League (1872,
and revived in 1881) that aligned Germany, Austria-Hungary, and Russia.  The
Triple Alliance was formed in 1882 that joined Germany, Austria-Hungary, and
Italy into a bloc that continued until World War I.  In 1890, a bipartite
agreement between Germany and Russia lapsed and this ultimately led to the
creation of a French-Russian alliance over the period 1891-94.  Subsequently
an Entente Cordiale between France and Great Britain was consummated in 1904,
and then a British-Russian agreement in 1907, that then bound France, Great
Britain, and Russia into the Triple Entente.  While our account of these
Byzantine maneuvers is incomplete (see Refs.~\cite{L} for more information),
and Fig.~\ref{ww1} does not show all relations and thus the extent of network
imbalance during the intermediate stages, the basic point is that these
relationship changes gradually led to a reorganization of the relations
between European nations into a socially balanced state.  Thus while social
balance is a natural outcome, it is not necessarily a good one!

\begin{figure}[htb]
\centerline{      \includegraphics[width=0.6\linewidth]{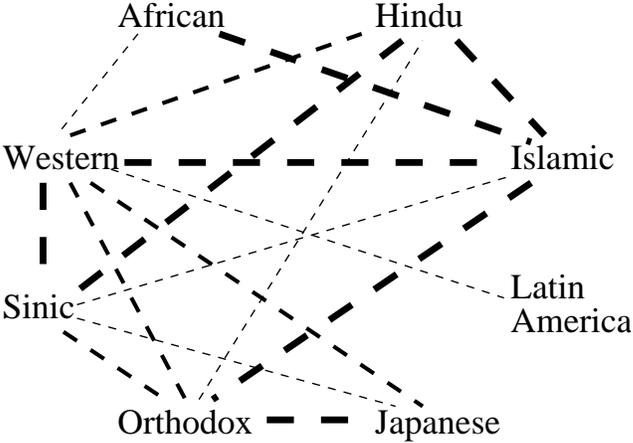} }
     \caption{Emerging conflicting relationships among major cultures as
       predicted by Huntington in 1996.  Thicker dashed lines represent
       stronger conflict.  This figure is a reproduction from the Wikipedia
       article on ``Clash of Civilizations'', Ref.~\cite{clash}. }
\label{Clash}
\end{figure}

Another more immediate, and perhaps more alarming, application of social
balance is to current international relations.  As popularized in
Huntington's book \cite{Hunt}, there appear to be developing civilizational
divisions across which increasing conflict is occurring (Fig.~\ref{Clash}).
According to Huntington, the division among humankind, and the source of
future conflict, will be predominantly cultural rather than ideological and
economic.  This thesis has generated a great deal of criticism, yet the core
idea---namely, that division and conflict is a more likely outcome rather
than the Westernized world's hope for a utopia because of global
democratization---may prove correct at least in the foreseeable future.

We close with some potentially interesting open theoretical questions.
First, it is natural consider more general interactions.  One can easily
imagine ternary relationships of friendly $+$, unfriendly $-$, or indifferent
$0$.  Another possibility is continuous-valued interaction strengths.  What
is the number of cliques and number of communities as a function of network
size and the density of indifferent relationships? Another direction, already
considered by Davis \cite{Davis}, is a more Machiavellian society in which
triads with three unfriendly relations are acceptable---that is ``an enemy of
my enemy may still be my enemy.''  This more relaxed definition for
imbalanced triads may lead to interesting dynamical behavior that will be
worthwhile to explore.  Finally, what happens if relations are not symmetric,
that is, $s_{ij} \ne s_{ji}$?  How does one define balance or some other
notion of social stability with asymmetric interactions?

TA gratefully acknowledges financial support from the Swiss National Science
Foundation under the fellowship 8220-067591.  SR acknowledges financial
support from NSF grant DMR0535503.


\begin{thebibliography}{99}
  
\bibitem{FH} F.~Heider, Psychol.\ Rev.\ {\bf 51}, 358--374 (1944); F.~Heider,
  J. Psychology {\bf 21}, 107--112 (1946); F.~Heider, {\em The Psychology of
    Interpersonal Relations} (J.~Wiley \& Sons, New York, 1958).
  
\bibitem{Lewin} K.~Lewin, {\em Field Theory in Social Science} (Harper, New
  York, 1951).
  
\bibitem{Newcomb} T.~M.~Newcomb, {\em The Acquaintance Process} (Holt,
  Rinehart \& Winston, New York, 1961).
  
\bibitem{social} S.~Wasserman and K.~Faust, {\em Social Network Analysis:
    Methods and Applications} (Cambridge University Press, New York, 1994).
  
\bibitem{CH} D.~Cartwright and F.~Harary, Psychol.\ Rev.\ {\bf 63}, 277--293
  (1956); F.~Harary, R.~Z.~Norman and D.~Cartwright, {\em Structural Models:
    An Introduction to the Theory of Directed Graphs} (John Wiley \& Sons,
  New York, 1965).
  
\bibitem{New} T.~M.~Newcomb,
  Social Psych.\ Quart.\ {\bf 42}, 299--506 (1979).

\bibitem{HNC} F.~Harary, R.~Z.~Norman and D.~Cartwright, {\em Structural
    Models: An Introduction to the Theory of Directed Graphs} (John Wiley \&
  Sons, New York, 1965).


\bibitem{Leik} R.~K.~Leik and B.~F.~Meeker, {\em Mathematical Sociology}
   (Prentice-Hall, Englewood Cliffs, N. J., 1975).
 
 \bibitem{B} P.~Bonacich, {\em Introduction to Mathematical Sociology}
   (http://www.sscnet.ucla.edu/soc/faculty/bonacich).
  
\bibitem{Davis} J.~A.~Davis, Human Relations {\bf 20}, 181--187 (1967).

\bibitem{Hum} N.~P.~Hummon and T.~J.~Fararo, J. Math.\ Sociology {\bf 20},
  145--159 (1995).
  
\bibitem{Dor} P.~Doreian and D.~Krackhard, J. Math.\ Sociology {\bf 25},
  43--67 (2001).
   
\bibitem{HD} N.~P.~Hummon and P.~Doreian, Social Networks {\bf 25}, 17--49
  (2003).
   
\bibitem{AKR} More details of our approach are given in T. Antal, P. L.
  Krapivsky, and S. Redner, Phys.\ Rev.\ E {\bf 72}, 036121 (2005).

\bibitem{K} A study of a similar spirit to ours is given in K. Kulakowski, P.
  Gawronski, and P. Gronek, Int.\ J. Mod.\ Phys.\ C {\bf 16}, 707 (2005);
  P.~Gawronski, P.~Gronek, and K.~Kulakowski, Acta Physica Polonica B {\bf
    36}, 2549 (2005).

\bibitem{fpp} We use the fact that the first-passage time to an absorbing
  point in a finite one-dimensional interval of length L with a bias away
  from the absorbing point is of the order of $e^{vL/D}$.  See S. Redner,
  {\it A Guide to First-Passage Processes}, (Cambridge University Press, New
  York, 2001).

\bibitem{vK}
  N.~G.~Van Kampen, {\em Stochastic Processes in Physics and Chemistry}
  (North Holland, Amsterdam, 2003).

\bibitem{Moore} See {\it e.g.}, M.~Moore, Eur.\ J.\ Social Psychology {\bf
    9}, 323--326 (1979).

\bibitem{L} W.~L.~Langer, {\em European Alliances and Alignments 1871--1890}
  (Knopf, New York, 1950, 2nd ed); B.~R.~Schmitt, {\em Triple Alliance and
    Triple Entente} (Holt, Rinehart, and Winston, Inc., New York, 1934).

\bibitem{clash} http://en.wikipedia.org/wiki/Clash$\underline{~}$of$\underline{~}$civilizations

\bibitem{Hunt}S. P. Huntington, {\em The Clash of Civilizations and the
    Remaking of World Order}, (Simon \& Schuster, New York, 1996); see also
  L. Harris,{\em Civilization and Its Enemies: The Next Stage of History},
  (The Free Press, New York, 2004).




\end{thebibliography}
\end{document}